\documentclass[smallextended]{svjour3}

\usepackage{amsmath}
\usepackage{amsfonts}
\spnewtheorem{assumptions}{Assumptions}{\bf}{\it}

\smartqed

\journalname{General Relativity and Gravitation}

\title{The interior solution of axially symmetric, stationary and rigidly rotating dust configurations}
\titlerunning{The interior solution of dust configurations}

\author{Norman G{\"u}rlebeck}

\date{Received: date / Accepted: date}

\institute{Norman G{\"u}rlebeck \at Institute of Theoretical Physics, Charles University, V Hole\v{s}ovick\'ach 2,\\180 00 Praha 8 - Hole\v{s}ovice, Czech Republic\\\email{norman.guerlebeck@gmail.com}}

\begin{document}

\maketitle

\begin{abstract}
It is shown that the interior solution of axially symmetric, stationary and rigidly rotating
dust configurations is completely determined by the mass
density along the axis of rotation. The particularly
interesting case of a mass density, which is cylindrical symmetric in the
interior of the dust configuration, is presented. Among other things, this proves the
non-existence of homogeneous dust configurations.
\end{abstract}

\section{Introduction}
Dust configurations played an important role in the search for global and physically meaningful solutions of
Einstein's field equations. Already in the 1920s, Lanczos obtained solutions
describing rigidly rotating, cylindrically symmetric and stationary dust configurations, see \cite{Lanczos_1924,Lanczos_1997}. A larger class of solutions including the one given by Lanczos was obtained by van Stockum in \cite{vanStockum_1937}. These solutions describe the interior of all rigidly rotating, axially symmetric and stationary dust configurations in terms of an arbitrary solution of a certain
second order partial differential equation. A closed form of these solutions involving one arbitrary function is given up to integrations in \cite{Islam_1983}.
Unfortunately, this arbitrary function lacks a direct physical interpretation.
One intention of this paper is to describe the degrees of freedom in the solutions
by a physically interpretable function, more precisely, the mass density given on
the axis of rotation.

A further generalization was made by Winicour in \cite{Winicour_1975}, where differential rotation was considered as well. In this case two functions can be chosen. One is completely arbitrary and the other must be an element of the kernel of the Laplacian in the flat three dimensional Euclidean space $\mathbb{R}^3$. However, in the present paper the attention is turned to rigidly rotating dust, i.e., the van Stockum class. 

In Newton's theory of gravity it was shown that \emph{isolated}\footnote{The support of the mass density is assumed to be compact.}, axially symmetric, rotating dust configuration unavoidably
collapse to a disk lying in a plane perpendicular to the axis of rotation and hence they cannot be stationary, see \cite{SchaudtPfister_2001} and references therein. However, this scenario can be prevented by distant, stabilizing matter distributions \cite{SchaudtPfister_2001}. Therefore, no assumptions about the exterior of the dust configuration are made in the approach presented here.

In general relativity a similar non-existence theorem is still lacking. But some partial results are already known. For instance, in \cite{Caporali_1978} it is shown that axially symmetric and stationary dust configurations do not yield an asymptotically flat spacetime provided that the mass density is strictly positive in the entire spacetime. 
This is also the reason, why the singularity in the mass density of Bonnor's dust cloud \cite{Bonnor_1977}, which is a special member of the van Stockum class, is inevitable. In this paper this result is generalized to dust configurations with a boundary seperating it from an arbitrary exterior, but the mass density is required to vanish on the boundary, see \cite{Zingg_etal_2007} as well. Furthermore, in \cite{Frauendiener_1987} the non-existence of dust configuration in a spatially compact manifold was proven. But note that in both theories, i.e., Newton's theory of gravity and general relativity, axially symmetric, stationary and rotating disks of dust perpendicular to the axis of rotation exist. An important example is given by the Neugebauer-Meinel disk of rigidly rotating dust and its Newtonian limit (Maclaurin disks) \cite{NeugebauerMeinel_1995}.

Bonnor's dust cloud serves as an example providing another interesting property of dust configurations in general relativity. The mass density of this solution admits a non vanishing gradient along the axis of rotation. This is not possible in Newton's theory of gravity, because there the angular velocity profile determines the mass density completely. In the case of rigid rotation this yields constant mass density. However, we will show that in general relativity there is no restriction on the mass density along the axis of rotation. In order to do so the Einstein equations for axially symmetric, stationary and rigidly rotating dust are solved for an arbitrary real analytic mass density given along the axis of rotation.
This mass density already yields the interior solution up to constants. Hence, the radial profile of the mass density is obtained, too. This can be
interesting for astrophysical observations. Moreover, it is even shown that dust configurations with the Newtonian mass density, i.e., homogeneous, do not exist in general relativity. The sole mass density constant
along the axis of rotation turns out to be the one given by Lanczos in the cylindrical
symmetric case \cite{Lanczos_1924}.

The paper is organized as follows. Section \ref{sec:Newton} is devoted to the Newtonian case. In particular, attention is paid to the interior solution and its implications.
It is shown that given the angular velocity curve the mass density is uniquely
determined and the gravitational potential up to an additive constant. After the
formulation of the problem of rigidly rotating, axially symmetric and stationary
dust configurations in general relativity in Sect. \ref{sec:GR} some non-existence
results for dust configurations with a boundary are proven in this framework. Afterwards, in Sect. \ref{sec:solution}, the solution of the interior field equations is obtained in terms of the mass density on the axis
of rotation. These results are used in Sect. \ref{sec:nonexistence} for a discussion of mass
densities, which are constant along the axis of rotation.

Throughout the text geometrical units, in which $c=G=1$ holds, are chosen.

\section{Dust configurations in Newton's theory of gravity}\label{sec:Newton}

To compare dust configurations in general relativity with the corresponding configurations in Newton's
theory of gravity some results regarding  stationary, axially symmetric rotating dust in the latter theory are recapitulated in this section.
The matter is characterized by a velocity field $\mathbf v$, which vanishes on the axis of 
symmetry, and a mass density $\mu$, such that
\begin{align}\label{velocity_massdensity}
 \mathbf v(\mathbf x,t)= \omega(\rho,\zeta)\rho \mathbf{e}_\varphi,\quad \mu(\mathbf x,t)=\mu(\rho,\zeta)
\end{align}
holds, where  $\omega=\omega(\rho,\zeta)$ denotes the angular velocity and $\{\rho,\zeta,\varphi\}$ are the cylindrical coordinates with the corresponding unit vectors $\{\mathbf e_{\rho},\mathbf e_{\zeta},\mathbf e_{\varphi}\}$ .
The gravitational potential $U$ has to satisfy the Poisson equation and the Euler equation
\begin{align}\label{Laplace_Euler}
   \nabla^2 U=4\pi \mu,\quad \mu\frac{\mathrm d}{\mathrm{d}t}\mathbf v=-\mu \nabla U.
\end{align}
The field equations and their consequences are investigated in an arbitrary open subset $\Omega$ 
of the support of the mass density. Hence, it is not necessary to exclude other matter distributions 
in the exterior of $\Omega$. Furthermore, the behavior at infinity is not restricted and cylindrically symmetric dust configurations are included in our considerations as well as distant objects stabilizing the dust configuration like in \cite{SchaudtPfister_2001}.

In this section we impose the following smoothness conditions for $\Omega$ and the functions $U$ and $\mu$:
\begin{assumptions}
\begin{enumerate} 
\item[]
\item The boundary of $\Omega$ denoted by $\partial\Omega$ is continuously differentiable,
\item the mass density $\mu\neq 0$ in $\Omega$,
\item $U\in C^2(\Omega)$.
\end{enumerate}
\end{assumptions}
Note that the first condition is only for concreteness. Others, not necessarily equivalent, like a Ljapunov surface \cite{Guenter} can be considered, too. The third condition leads with the Poisson equation \eqref{Laplace_Euler} to $\mu\in C^0(\Omega)$ and thus with the Euler equation to $\omega\in C^1(\Omega)$. Weaker differentiability conditions are possible, if weak solutions for the gravitational potential are considered as well.

From the Euler equation \eqref{Laplace_Euler} and the particular form of the velocity field 
\eqref{velocity_massdensity} it follows that
\begin{align}\label{Derivatives}
	U_{,\varphi}=U_{,\zeta}=0,\quad U_{,\rho}=\rho\omega(\rho,\zeta)^2,
\end{align}
where a comma denotes a partial derivative. Therefore, the gravitational potential $U$ is cylindrically symmetric in $\Omega$ and consequently the angular velocity. Hence, $U$ is given up to an additive constant in $\Omega$ by
\begin{align}\label{innersolution}
	U(\rho)=U_{0}+\int\rho\omega(\rho)^2\mathrm{d}\rho.
\end{align}
The constant $U_0$ can be fixed using the solution in the exterior of $\Omega$ and a smoothness condition for $U$ across $\partial\Omega$.
If \eqref{Derivatives} is inserted in the Poisson equation \eqref{Laplace_Euler} all possible mass densities can be obtained given an angular velocity profile by
\begin{align}\label{mass_density_vs_angular_velocity}
	(\omega(\rho)^2\rho)_{,\rho}+\omega(\rho)^2=4\pi\mu(\rho).
\end{align}
Therefore, for every angular velocity curve exists one and only one mass density. In the case of $\omega=\omega_0$ in $\Omega$, i.e., rigidly rotating dust, the constant mass density
\begin{align}\label{homogeneous}
	\mu=\frac{\omega_{0}^2}{2\pi}
\end{align}
is obtained. It is worth noting that this proves the non-existence of static dust configurations under the assumptions 1 in Newton's theory of gravity. With \eqref{innersolution} the solution is completely determined by the constant $\omega_0$ in $\Omega$.

Conversely, the angular velocity is determined for a given mass density as a solution of the ordinary differential equation \eqref{mass_density_vs_angular_velocity}, which is given by
\begin{align}
	\omega(\rho)=\pm\frac{1}{\rho}\sqrt{4\pi\int\mu(\rho')\rho'\mathrm{d}\rho'+\alpha}.
\end{align}
If the rotational axis intersects $\Omega$, the constant $\alpha$ must be chosen to preserve the differentiability of the angular velocity in $\Omega$. If it does not, additional information about the solution in the exterior of $\Omega$ is necessary in order to determine $\alpha$.

The non-existence of isolated, axially symmetric and stationary dust configurations in vacuum can be shown provided that $U\in C^1(\mathbb{R}^3)$ and $U$ vanishes in infinity, see \cite{Bonnor_1977,SchaudtPfister_2001}. 

\section{Rigidly rotating dust configurations in general relativity}\label{sec:GR}

In the case of general relativity we restrict ourselves to axially symmetric and stationary spacetimes. Hence, it is convenient to use the Lewis-Papapetrou line element in quasi cylindrical coordinates
\begin{align}\label{metric}
\mathrm{d}s^2=\mathrm{e}^{-2U}[\mathrm{e}^{2k}(\mathrm{d}\rho^2+\mathrm{d}\zeta^2)+W^2\mathrm{d}\varphi^2]-\mathrm{e}^{2U}(\mathrm{d}t+a\mathrm{d}\varphi)^2,
\end{align} 
where the functions $U,k,a,W$ depend only one the coordinates $\rho,~\zeta$. Since the field equations are discussed in a spacetime region $\mathcal{G}$, where the matter can be interpreted as dust,
the function $W$ can be chosen to be the radial coordinate $\rho$ by means of a conformal mapping. Furthermore, for rigidly rotating dust a transformation in a co-moving coordinate system is possible without changing the form of the metric. Let us for simplicity of notation assume the metric \eqref{metric} is already given in these co-moving, canonical Weyl-coordinates and let us denote with $\Omega$ an open subset of $\mathbb R ^3$, such that the closure of $\Omega$ is a subset of the restricted coordinate map of $\mathcal{G}$ with respect to $\{\rho,\varphi,\zeta\}$.

The only non-vanishing component of the stress-energy tensor in $\mathcal{G}$ reads
\begin{align}
	T^{tt}=\mu \mathrm{e}^{-2U}
\end{align}
in this coordinate system, where $\mu$ denotes the non-negative mass density.

In analogy to the last section we assume the following:
\begin{assumptions}\label{ass:2}
\begin{enumerate}
	\item[]
	\item The boundary $\partial\Omega$ of $\Omega$ is continuously differentiable\label{boundary},
	\item the mass density $\mu\neq 0$ in $\Omega$\label{massdensity}, 
	\item	$U,~a,~k\in C^2(\Omega)$.\label{differentiability}
\end{enumerate}
\end{assumptions} 
Note that again the first condition could be substituted by others like the Ljapunov conditions. Furthermore, it is not assumed that the dust configuration is an isolated object, i.e., that the spacetime is asymptotically flat. Other matter distributions can be present in the exterior of $\mathcal G$. In particular, if several non-connected components of the dust configuration exist, they can be treated independently in the approach to be described.

If we denote the part of the axis of rotation, which intersects $\Omega$, by $\mathcal{A}$, then it is convenient for the Theorem \ref{localformofsolution} to formulate a second set of assumptions:
\begin{assumptions}\label{ass:3}
\begin{enumerate}
	\item[]
	\item The set $\mathcal A$ is not empty,
	\item the origin $(\rho,\zeta)=(0,0)$ lies in $\mathcal{A}$,
	\item the spacetime is elementarily flat. 
\end{enumerate}
\end{assumptions} 
If the first condition holds, the second can always be realized by a coordinate shift in the $\zeta$-direction. If the Assumption \ref{ass:2}.\ref{differentiability} and the elementary flatness condition are satisfied, then
\begin{align}\label{ataxis}
 a_{,\rho},~k_{,\rho}~U_{,\rho}\in O(\rho)\quad\text{and}\quad a_{,\zeta},~k_{,\zeta}\in O(\rho^2)
\end{align}
holds.  

Now we turn our attention to the field equations. The contracted Bianchi identity $T^{ab}_{\phantom{ab};a}=0$ and the Assumptions \ref{ass:2} imply that the function $U$ must be a finite constant $U_0$ in $\Omega$. Therefore, the non-redundant field equations simplify in $\Omega$, see, e.g., \cite{stephani_etal_2003}, to
\begin{align}\label{untransformed_fieldequations}
	\frac{\mathrm{e}^{6U_0}}{\rho^2}\left(\nabla a\right)^2=8\pi\mu\mathrm{e}^{2k},\quad\quad \Delta a -\frac{a_{,\rho}}{\rho}=0,
\end{align}
where $\Delta$ is the Laplace operator in cylindrical coordinates in the three dimensional Euclidean space for axially symmetric functions. The function $k$ is given by the line integration
\begin{align}\label{lineintegration}
	k=\frac{\mathrm{e}^{4U_0}}{4}\int \left[\frac{1}{\rho}((a_{,\zeta})^2-(a_{,\rho})^2)\mathrm{d}\rho-\frac{2}{\rho} a_{,\rho}a_{,\zeta}\mathrm{d}\zeta\right].
\end{align}
Equations \eqref{untransformed_fieldequations} and \eqref{lineintegration} are well defined in a neighborhood of the rotation axis because of \eqref{ataxis}.  

It is a well known fact that the field equations in the vacuum can be simplified to the Ernst equations with the transformation
\begin{align}\label{Transformation auf b}
	b_{,\rho}=-\frac{a_{,\zeta}}{\rho}\mathrm{e}^{4U},\quad  	   	b_{,\zeta}=\frac{a_{,\rho}}{\rho}\mathrm{e}^{4U}.
\end{align}
The integrability condition of this transformation holds because of the field equations \eqref{untransformed_fieldequations} in $\Omega$, too, and the function $b$ is twice continuously differentiable in $\Omega$ because of Assumptions \ref{ass:2}. The transformed field equations read
\begin{subequations}\label{fieldequations}
\begin{align}
	\Delta b&=0,\label{fieldequation_b}\\
	\quad \left(\nabla b\right)^2&=8\pi\mu\mathrm{e}^{2k+2U_0}\label{fieldequation_b_mass},\\
	k&=\frac{\mathrm{e}^{-4U_0}}{4}\int\rho\left[((b_{,\rho})^2-(b_{,\zeta})^2)\mathrm{d}\rho+
	2b_{,\rho}b_{,\zeta}\mathrm{d}\zeta\right]\label{fieldequation_k},
\end{align}
\end{subequations}
where the first equation is the integrability condition of the inverse transformation.
The behavior of the functions $a,~k$ close to the axis \eqref{ataxis} ensures that the transformation \eqref{Transformation auf b} and the field equations \eqref{fieldequations} are also valid on the axis. Note that since $b$ is harmonic it is real analytic in $\Omega$, as well. With \eqref{fieldequation_k} and \eqref{fieldequation_b_mass} $k$ and $\mu$ are real analytic, too. Therefore, singularities in the mass density are excluded by the assumptions and the field equations. Conversely, only real analytic mass densities can be given in order to obtain a solution $b$ in $C^2(\Omega)$, which is more restrictive than in Newtonian physics, where only a continuous mass density is required.

\section{The solution of the field equations}\label{sec:solution}

Before the general solution is obtained, we prove some non-existence statements in the formalism presented in the last section.
\begin{theorem}\label{theorem:nonexistence1}
Let us suppose the Assumptions \ref{ass:2}.\ref{boundary} and \ref{ass:2}.\ref{massdensity} are satisfied. Then the field equations \eqref{fieldequations} do not admit solutions $b,k,U_0\in C^2(\Omega)\cap C(\Omega\cup\partial\Omega)$ in $\Omega$, if one of the following properties is satisfied:
\begin{enumerate}
	\item The mass density $\mu\in C^0(\Omega\cup\partial\Omega)$ vanishes on $\partial\Omega$\label{nonexistencemassdensity},
	\item $b$ is constant on $\partial\Omega$,\label{nonexistencetangential}
	\item the normal derivative of $b$ on $\partial\Omega$ vanishes.\label{nonexistencenormal}
\end{enumerate}
\end{theorem}
\begin{proof}
The first case can be reduced to the third using \eqref{fieldequation_b_mass}. The cases \ref{nonexistencetangential} and \ref{nonexistencenormal} follow directly from the uniqueness of the solution of the Laplace equation for $b$ under the given assumptions. In both cases the unique solution is given by $b=\mathrm{const}.$ which in return leads to $\mu=0$ in $\Omega$ with the field equation \eqref{fieldequation_b_mass} and the fact that $\mathrm{e}^{2k+2U}$ is finite and positive, which can be seen from the differentiability assumptions. However, this is in contradiction with the definition of $\Omega$ and the Assumption \ref{ass:2}.\ref{massdensity}.\qed
\end{proof}
The first part was also shown in \cite{Zingg_etal_2007}. The theorem includes also the fact that solutions in the van Stockum class, which describe a spacetime filled completely with dust and a mass density vanishing at infinity,\footnote{In order to interpret the last theorem physically the assumptions, which were necessary to ensure the validity of the transformation \eqref{Transformation auf b} and which were summarized in the last section, have to hold.} necessarily have to violate some of the assumptions of Theorem \ref{theorem:nonexistence1}, see, e.g,. for other proofs \cite{Caporali_1978} or for a recent approach \cite{Bratek_etal_2007}. In particular the
differentiability conditions of $b$ are not satisfied by the solution given in \cite{Bonnor_1977}, because there is a singularity at the origin.

As we will show in the remainder of this section the conditions which are implied by the assumption of rigid rotation are not as restrictive as in the case of Newton's theory of gravity. More precisely, to \emph{every} real analytic mass density chosen arbitrarily at $\mathcal A$ two solutions of the inner field equations in $\Omega$ can be assigned at least locally. Let us denote by $B_\epsilon$ the open ball with the radius $\epsilon$ and the origin of ${\mathbb R}^3$ as center.
 
\begin{theorem}\label{localformofsolution}
Let us suppose that Assumptions 2 and 3 hold. Furthermore, let us assume that the mass density $\mu$ is real analytic in $\zeta$ in a neighborhood of the origin with the radius of convergence $\epsilon$ of the series expansion in $\zeta$. Then the solution $b\in C^2(\Omega)$ of the field equations \eqref{fieldequations} is completely determined in a non-empty set $B_\sigma\subset B_\epsilon\cap\Omega$ by the mass density and its derivatives at the origin, an arbitrary constant $b_{\pm}(0)$ and a choice of a sign:
 \begin{align}\label{formofsolution}
	b_{\pm}=b_{\pm}(0)\pm\sqrt{8\pi{\mathrm e}^{2U_0}}\sum\limits_{l=1}^{\infty}
	\frac{1}{l!}(\sqrt{\mu})^{(l-1)}r^lP_l(\cos\theta),
\end{align}
where $(\sqrt{\mu})^{(n)}$ denotes the $n$th derivative of $\sqrt{\mu}$ with respect to $\zeta$ at the
point $(\rho,\zeta)=(0,0)$. The $P_l$ denote the Legendre polynomials of the first kind. Furthermore, polar coordinates $\rho=r\sin\theta$ and $\zeta=r\cos\theta$ are used.
\end{theorem}
\begin{proof}
	With \eqref{ataxis} and \eqref{Transformation auf b} it follows that $b_{,\rho}$ vanishes along $\mathcal A$.
	Hence, the second field equation \eqref{fieldequation_b_mass} simplifies along $\mathcal A$ to
	\begin{align}\label{bonaxismu}
		b_{,\zeta}=\pm\sqrt{8\pi\mu{\mathrm e}^{2U_0}}.		
	\end{align}
	Therefore, the mass density given along $\mathcal{A}$ determines the function $b$ up to a sign 
    and a constant. Since $\Omega$ is an open set and $(\rho,\zeta)=(0,0)$ is assumed to be an inner 
	point a radius $\sigma>0$ exists, such that $B_\sigma\subset B_\epsilon\cap\Omega$. In $B_\sigma$  
	the square root of the mass density at the axis admits the convergent series expansion
	\begin{align}\label{seriesexpansionmu}
		\sqrt{\mu(0,\zeta)}=\sum_{l=0}\frac{1}{l!}(\sqrt{\mu})^{(l-1)}\zeta^l.
	\end{align}
  Because $b$ is an axially symmetric harmonic function in $B_{\sigma}$, it can be written in the form
	\begin{align}\label{seriesexpansionb}
		b(r,\theta)=\sum_{l=0}^{\infty}A_lr^lP_l(\cos\theta).
	\end{align}
The coefficients $A_l$ can be
	derived using the identity theorem of power series, \eqref{bonaxismu},	\eqref{seriesexpansionmu} and
	\eqref{seriesexpansionb}. This yields the coefficients $A_l$ given in \eqref{formofsolution}.\qed
\end{proof}
Some remarks are expedient here. The introduction of the set $B_{\sigma}$ is for purely technical reasons, i.e., to avoid different assumptions, e.g about the topology of $\Omega$. If the convergence radius $\epsilon$ is such that $B_\epsilon\supset\Omega$ and $\Omega$ is a region, then the result \eqref{formofsolution} can be extended to the entire set $\Omega$.
The assumptions about the analyticity of $\mu$ is necessary and sufficient in order to obtain a solution $b$ of the field equations in accordance with the differentiability assumptions.

To obtain also solutions admitting a non-constant mass density along the axis seems surprising in the light of the results in Newtonian gravity in Sect. 2, where a constant mass density is implied by rigid rotation \eqref{homogeneous}. One possible explanation is that gravitomagnetic effects due to the motion of the dust will act like a force in $\zeta$-direction. Thus, other solutions of a generalized Euler equation \eqref{Laplace_Euler}, at least in a slow motion limit, are possible. These results and how to assign to such solutions a proper Newtonian limit using Ehlers frame theory will be discussed elsewhere.

The constants in \eqref{formofsolution} cannot be determined any further. If $b$ is a solution of the field equations \eqref{fieldequations} in $\Omega$ so are $b+\mathrm{const}.$ and $-b$ and the same mass density is obtained from them. The constant $b_{\pm}(0)$ is the usual freedom due to the transformation formulas \eqref{Transformation auf b}.

Theorem \ref{localformofsolution} provides us together with the field equation \eqref{fieldequation_k}
with an algorithm to determine the general solution of the field equation if an 
analytic mass density is given along the axis. By \eqref{formofsolution} $b$ is obtained from the mass 
density up to a constant and a sign. With \eqref{fieldequations} the function $k$ as well as the mass density can 
be determined independently of the chosen constant and sign. The sign in \eqref{formofsolution} and the constant $U_0$ can be fixed, provided that a solution of the field equations in the exterior of $\Omega$ is known. 

In order to obtain an exterior solution of the Einstein equations several approaches are possible. In some cases the mass density along the axis can be extended in $\zeta$, e.g., if the radius of convergence of the Taylor expansion in $\zeta$ is infinite such that a globally valid cosmological solution can be obtained and no exterior solution arises. Such spacetimes describe a universe filled with axially symmetric, stationary, rigidly rotating dust.

If the interior solution described in \eqref{formofsolution} should be joined to an asymptotically flat vacuum exterior one has to solve a ``Dirichlet problem'' with a free boundary for the Ernst equation. Whether such a solution exists, especially in the light of the non-existence of such dust configurations in the Newtonian gravity, is still an open and difficult question. Perhaps our form of the interior solution will prove useful to answer it. However, if the vacuum exterior is \emph{not} supposed to be asymptotic flat, then global solutions exist, e.g., van Stockums cylindrically symmetric dust \cite{vanStockum_1937} or \cite{Vishveshwara_1977} joining the dust to a vacuum exterior. 

Another possibility would be to consider an exterior, where matter can be present. A first approach to such an ``stabilizing'' matter configuration could be to consider a shell enclosing the dust configuration. However, one has to solve a ``Dirichlet problem'' for the Ernst equation in the region between the dust and the shell and a ``Dirichlet problem'' for the asymptotic flat vacuum region outside the shell. Even though this problem is not trivial the limiting case of a shell situated on the surface of the dust seems feasible. These ``dust stars with a crust'' will be investigated in future work.

\section{The non-existence of homogeneous dust configurations}\label{sec:nonexistence}

The algorithm described above is now applied to the important example of constant mass density along the rotation axis. It was shown in Sect. \ref{sec:Newton} that this was the sole possible case for rotating Newtonian dust configurations \eqref{Derivatives} and that in the case of rigid rotation the mass density must be homogeneous \eqref{homogeneous} in $\Omega$. As is proven in the following corollary this does not hold in general relativity. The solutions of the field equations in $\Omega$ for mass densities independent of the $\zeta$ coordinate do \emph{not} yield a homogeneous mass density.

\begin{corollary}
If a dust configuration satisfies Assumptions \ref{ass:2} and \ref{ass:3} and the mass density $\mu$ is constant $\mu=\mu_0\neq0$ along the axis of symmetry, then there exists a $\sigma>0$ such that the mass density is given in $B_\sigma\subset \Omega$ by
\begin{align}\label{massdensity_const}
		\mu(\rho,\zeta)=\mu_0{\mathrm{exp}}(2\pi\mu_0{\mathrm e}^{-2U_0}\rho^2).
\end{align}
\end{corollary}
\begin{proof}
Because the mass density is constant in $\mathcal A$ all derivatives with respect to $\zeta$ vanish at the origin and the convergence radius $\epsilon$ of the series representation at this point is infinite. Using the Theorem \ref{localformofsolution} and the transformation between cylindrical and polar coordinates the solution of the field equations \eqref{fieldequation_b} and \eqref{fieldequation_b_mass} can be written as
\begin{align}\label{solution_b_const}
	b(\rho,\zeta)=b_{\pm}(0)\pm\sqrt{8\pi{\mathrm e}^{2U_0}\mu_0}\zeta
\end{align}
in a $B_\sigma\subset\Omega$ with $\sigma>0$. The function $k$ is obtained by means of the line integration \eqref{fieldequation_k} and \eqref{ataxis} 
\begin{align}\label{solution_k_const}
	k=-\pi\mu_0\mathrm e^{-2U_0}\rho^2.
\end{align}
Inserting \eqref{solution_b_const} and  \eqref{solution_k_const} in \eqref{fieldequation_b} yields the mass density given in \eqref{massdensity_const} in $B_\sigma$.\qed
\end{proof}

This corollary does not only prove the non-existence of homogeneous dust configurations, it also gives the only possible mass density in the cylindrically symmetric case as obtained by Lanczos, see \cite{Lanczos_1924}. But here $\Omega$ need not be cylindrically symmetric. Only $\mu$ must be independent of $\zeta$ on the axis of rotation.

\begin{acknowledgements}
I thank R. Meinel for turning my attention to this interesting topic. Fertile discussions with R. Meinel and D. Petroff are also gratefully acknowledged. 
\end{acknowledgements}

\end{document}